\documentclass[twocolumn]{aastex63}
\usepackage{amsmath}
\usepackage{bm}

\received{today}
\submitjournal{ApJ}

\shorttitle{Multi-Wave band Polarization of GRB Afterglows}
\shortauthors{Shimoda \& Toma}

\begin{document}

\title{Multi-Wave band Synchrotron Polarization of Gamma-Ray Burst Afterglows}

\correspondingauthor{Jiro Shimoda}
\email{shimoda.jiro@k.mbox.nagoya-u.ac.jp}

\author[0000-0003-3383-2279]{Jiro Shimoda}
\affiliation{Frontier Research Institute for Interdisciplinary Sciences,
Tohoku University, Sendai 980-8578, Japan}
\affiliation{Astronomical Institute, Graduate School of Science, Tohoku University, Sendai 980-8578, Japan}
\affiliation{Department of Physics, Graduate School of Science, Nagoya University, \\
Furo-cho, Chikusa-ku, Nagoya 464-8602, Japan}

\author[0000-0002-7114-6010]{Kenji Toma}
\affiliation{Frontier Research Institute for Interdisciplinary Sciences,
Tohoku University, Sendai 980-8578, Japan}
\affiliation{Astronomical Institute, Graduate School of Science, Tohoku University, Sendai 980-8578, Japan}




\begin{abstract}
Multi-wave band synchrotron linear polarization of gamma-ray burst (GRB)
afterglows is studied under the assumption of an anisotropic turbulent magnetic
field with a coherence length of the plasma skin-depth scale in the downstream of
forward shocks. We find that for typical GRBs, in comparison to the optical
polarization, the degree of radio polarization shows a similar temporal evolution
but a significantly smaller peak value.
This results from differences
in observed intensity image shapes between the radio and optical bands. We also
show that the degree of the polarization spectrum undergoes a gradual variation from
the low- to the high-polarization regime above the intensity of the spectral peak
frequency, and that
the difference in polarization angles in the two regimes is zero or 90 degrees.
Thus, simultaneous multi-wave band polarimetric observations of GRB afterglows
would be a new determinative test of the plasma-scale magnetic field model.
We also discuss
theoretical implications from the recent detection of radio linear polarization
in GRB 171205A with ALMA and other models of magnetic field configuration.
\end{abstract}



\section{Introduction}
\label{sec:intro}
Gamma-ray bursts (GRBs) are the brightest transients in the gamma-ray sky. The
progenitor of some long-duration GRBs was established as the core collapse of massive
stars \citep[and references therein]{hjorth12}, and that of one short-duration
GRB as coalescence of a double neutron star binary \citep{abbott17}. Extensive
efforts in lightcurve and spectral observations of GRBs and theoretical arguments
led to the standard scenario: The progenitor system gives rise to a new-born
black hole with an accretion disk, and drives a relativistic jet, which emits
prompt gamma-rays. Then the interaction of the jet with the circumburst medium
generates a reverse shock and a forward shock, which emit long-lived afterglows
observed in the wave bands from radio to gamma-rays. In particular, synchrotron
emission of nonthermal electrons created by the forward shocks can explain
the observational results of many late-phase afterglows \citep[for reviews]{meszaros02,
piran04,kumar15}.
\par
Several major problems remain concerning the standard scenario, however,
which include the origins of the strong magnetic field and the
nonthermal electrons in the shocked region \citep[][for a review]{sironi15}.
Relativistic collisionless shocks are formed with amplification of the
magnetic field on the plasma skin-depth scale by Weibel instability,
although whether the amplified magnetic field can be maintained at a sufficiently long distance
in the downstream of a shock \citep{medvedev99,gruzinov99,keshet09,sironi11,asano20}
is a subject of debate. Other
mechanisms of field amplification on larger scales have also been investigated
\citep{sironi07,inoue13,duffell14,tomita19}. Particle-in-cell simulations of relativistic
shocks show nonthermal electron production \citep{sironi13}, but the physical mechanism of electron
acceleration is still elusive since the electron spectral index and maximum energy
depend on the unknown magnetic field properties. Furthermore, the number ratio of
nonthermal to thermal electrons has not been probed by observations yet \citep{toma08},
and this affects the estimate of the total energies of GRBs \citep{eichler05}, which then
affects the estimate of the detectability of GRBs' multi-messenger and high-redshift signals
\citep{murase06,kimura17,s_inoue13,toma16}.
\par
Polarimetric observations in addition to the light-curve and spectral ones are powerful
for solving such problems on collisionless shock physics \citep[for reviews][]{lazzati06,
toma13,covino16}. During the last decade, polarimetric observations have achieved a number
of new measurements and indications; in gamma-ray prompt emission with GAP \citep{yonetoku11,
yonetoku12}, INTEGRAL \citep{gotz09,gotz13}, AstroSat \citep{sharma19}
and POLAR \citep{zhang19,kole19,burgess19},\footnote{Reports of
gamma-ray polarization measurements include those of GRB 021206 by RHESSI \citep{coburn03}
which were refuted by \citet{rutledge04} and \citet{wigger04} and those of GRB 041219A by
INTEGRAL SPI \citep{kalemci07} were inconsistent with the same event data by INTEGRAL IBIS
\citep{gotz09}. We should note that even for the recent measurements their detection
significances are mostly $\sim 3\sigma$.} in optical prompt emission and early afterglows
with the Liverpool Telescope \citep{mundell13,kopac15, steele17,jordana20}, the Kanata Telescope
\citep{uehara12}, and Master Telescopes \citep{troja17}, and in radio afterglows with the ALMA
\citep{urata19,laskar19}. Circular polarization has also been detected
in the optical band \citep{wiersema14}. 
\par
The radio polarization had not been detected until very recently, probably because
the observed spectral parts ($\simeq 8.4\;$GHz with the VLA in most cases) were optically
thick due to synchrotron self-absorption \citep{taylor05,granot05,granot14,vanderhorst14}.
The recent two detections were performed for optically thin parts at the higher
frequency with ALMA, $\simeq 97.5\;$GHz. One of them revealed that the polarization
degree (PD) in the forward shock afterglow of GRB 171205A, $\Pi \simeq 0.27\%
\pm 0.04\%$~\citep{urata19}\footnote{This detection is now under debate. \citet{laskar20}
claimed that the systematic error of polarimetric calibration is $\sim 0.1~\%$, which
is more than three times larger than the value in the ALMA technical handbook, and then
inferred a 3$\sigma$ upper limit $\Pi < 0.30\%$.}, is significantly lower than the typical
late-phase optical afterglow PD, $\Pi \sim 1-3\%$ \citep{covino04,covino16,urata19,stringer20}. 
\par
The difference between the optical and radio polarizations was theoretically discussed by
focusing on the Faraday rotation effect by thermal electrons or radiatively-cooled electrons in the
shocked region with large-scale magnetic turbulence \citep{matsumiya03,sagiv04,toma08}.
For the anisotropic turbulent magnetic field with a coherence length on
the plasma skin-depth scale
\citep{medvedev99,sari99,ghisellini99,rossi04,gill20}, the Faraday rotation effects are tiny, and
little attention has been paid to the wavelength dependence of polarization \citep[but see][]{mao17}. Only \citet{rossi04}
showed a calculation result in the plasma-scale turbulent field model
with PD at the observed
frequency of $\nu \ll \nu_{\rm m}$ (which they refer to as the `radio branch')
is lower than that at
$\nu \gg \nu_{\rm m}$ (`optical branch'), where $\nu_{\rm m}$ is the spectral peak frequency of
afterglow synchrotron emission. However, they did not clarify the physical reason for
this result
nor did they mention on how PD varies around $\nu \sim \nu_{\rm m}$, for which ALMA polarimetric
observations will be often performed, as suggested by GRB 171205A \citep{urata19}.
\par
In this paper, we calculate the PD temporal variations of GRB afterglows at multiple
wave bands
in the anisotropic plasma-scale turbulent magnetic field model, and pin down the physical reason for the
wavelength dependence of PD by showing the relationship of the afterglow image on the sky with
the net PD when observed as a point source. Furthermore, we derive PD spectra, in which we find
gradual PD variations above $\nu \sim \nu_{\rm m}$, and show their parameter dependence. These
analyses will be crucial for testing the plasma-scale magnetic field model with polarimetric data
at ALMA bands (and simultaneous optical polarimetric data) and be useful for theoretical prediction
of radio polarization for different magnetic field models.
\par
This paper is organized as follows. In Section~\ref{sec:jet}, we introduce a standard model of
the dynamics and emission fluxes of the expanding forward shocks. Then, in Section~\ref{sec:synch}, we
consider synchrotron emission and its polarization with an assumption of an anisotropic plasma-scale
turbulent magnetic field. After explaining the parameter sets that we use in Section~\ref{sec:parameter},
we show the calculation results of multi-wave band synchrotron polarization and their physical
interpretation in Section~\ref{sec:results}. Section~\ref{sec:discussion} is devoted to a summary
of our findings and a discussion.

\section{Blast wave dynamics and emission flux}
\label{sec:jet}
\begin{figure}
\includegraphics[scale=0.6]{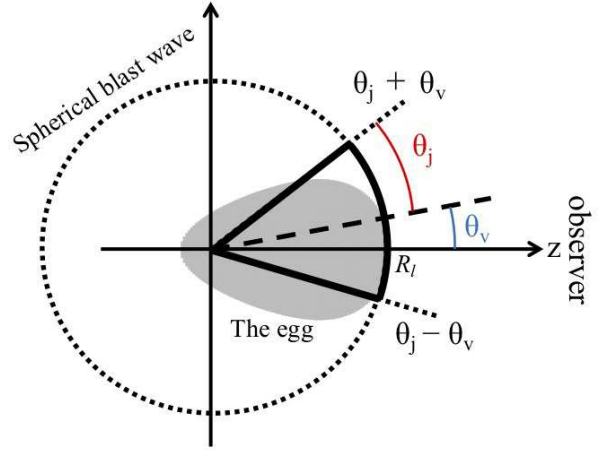}
\caption{Schematic diagram of a collimated blast wave.
The dotted curve corresponds to the shock surface of
a spherical blast wave, which expands radially. We
only consider the part of the blast wave within the angle
interval $2\theta_{\rm j}$ (thick solid line) as produced
by a GRB jet. Our line of sight is fixed along the
$z$-axis, which makes an angle $\theta_{\rm v}$ with the
jet axis (dashed line). For the spherical blast wave, photons
from the shaded region (egg) have the same arrival time $T$.
The collimated region of the egg is also described by Figure~\ref{fig:egg},
in which the white (black) line indicates $\theta=\theta_{\rm j}
-\theta_{\rm v}$ ($\theta=\theta_{\rm j}+\theta_{\rm v}$)
for $\phi=0$ and $\pi$.
}
\label{fig:jet}
\end{figure}

We consider GRB afterglows as synchrotron emission from relativistically expanding
spherical blast waves, and calculate their dynamics and emission fluxes by following
the formulation of \cite{granot99} and taking into account of the collimation of outflows.
The radius of the shock front is $R = R(t)$ and its Lorentz factor ${\it \Gamma}$
is proportional to $R^{-3/2}$ (adiabatic expansion, i.e., constant
explosion energy).  The internal structure of the
shocked region with ${\it \Gamma}\gg1$
is given by~\citep[cf.][eqs. 28-30 and 40-42]{blandford76}
%
\begin{eqnarray}
n' &=& 4\gamma_f n_0 \chi^{-\frac{5}{4}}, \\
\gamma &=& \gamma_f \chi^{-\frac{1}{2}}, \\
e' &=& 4 n_0 m_{\rm p} c^2 \gamma_f{}^2\chi^{-\frac{17}{12}},
\label{eq:BM}
\end{eqnarray}
%
where the superscript prime denotes a value measured in the rest frame of the fluid, $n$
is the number density, $\gamma$ is the Lorentz factor of the fluid, and $e$ is the internal
energy, respectively. $\gamma_f\simeq{\it \Gamma}/\sqrt{2}$ indicates the Lorentz factor
of the fluid just behind the shock. The ambient number density measured
in the observer frame is $n_0$.  The speed of light and proton mass are $c$ and
$m_{\rm p}$, respectively. The self-similar variable $\chi$ is defined as
%
\begin{eqnarray}
\chi = 1 + 16\gamma_f{}^2\left(\frac{R-r}{R}\right).
\end{eqnarray}
%
Note that the shocked region may have a thickness of $\Delta\sim R/4\gamma^2$. The radius
$R$ can be rewritten as function of a photon arrival time to the observer, $T$. We use a
spherical coordinate system centered on the blast wave and set our line of sight along the
$z$-axis for convenience. For a photon emitted at time $t$ and position $(r,\mu)$ in the
observer frame, where $\mu\equiv\cos\theta$, the arrival time is
%
\begin{eqnarray}
T_z=\frac{T}{1+z}=t-\frac{r\mu}{c},
\label{eq:arrival time}
\end{eqnarray}
%
where $z$ is the cosmological redshift and we choose that $T=0$ as the arrival time of a
photon emitted from the origin at time $t=0$. Solving the motion equation of the shock,
${\rm d}R/c{\rm d}t = \sqrt{1-1/{\it \Gamma}^2}$ with ${\it \Gamma}\propto R^{-3/2}$ and
${\it \Gamma}\gg1$, we obtain
%
\begin{eqnarray}
  R\simeq\frac{cT_z}{1-\mu+1/(8{\it \Gamma}^2)}.
\label{eq:egg}
\end{eqnarray}
%
The surface described by $R=R(\mu,T)$ at which the photons have the same arrival time
forms an elongated `egg' shape that is no longer spherically symmetric, as illustrated
in Figure~\ref{fig:jet}. Such a relativistic lookback geometry was
first introduced for constant $\Gamma$ by~\citet{rees66}. Equation~(\ref{eq:egg}) is a
version of the geometry for $\Gamma \propto R^{-3/2}$. We will refer to this surface
as `the egg' \citep[cf.][]{granot99}. $\chi$ is now a self-similar variable dependent
on $T$, $r$, and $\mu$.
%
%
\par
It is useful to additionally introduce another self-similar variable,
%
\begin{eqnarray}
y=\frac{R}{R_l},
\end{eqnarray}
%
where $R_l$ is the semi-major axis of the egg. Then we can describe the condition
of adiabatic expansion as
%
\begin{eqnarray}
\gamma_f=\gamma_l y^{-\frac{3}{2}},
\end{eqnarray}
%
where $\gamma_l$ is the Lorentz factor of the fluid just behind the shock at the semi-major
axis of the egg. From Eq.~\eqref{eq:arrival time}, we find that relationships between
the variables $(r,\mu)$ and $(\chi,y)$ are
%
\begin{eqnarray}
r \simeq R_l y, ~~~~~
\mu \simeq 1 - \frac{ 1-\chi y^4 }{16\gamma_l{}^2 y}.
\end{eqnarray}
%
Note that the position $(\chi,y)=(1,1)$ presents $(r,\mu)=(R_l,1)$, and $R_l=8{\it \Gamma}^2
c T_z=16\gamma_l{}^2 cT_z$.
\par
\par
Figure~\ref{fig:jet} schematically shows the geometry of our system.
We regard a part of the blast wave within the angle interval $2\theta_{\rm j}$
as produced by a GRB jet, and exclude the other part. The viewing angle
$\theta_{\rm v}$ is defined as the angle between the jet axis and our
line of sight ($z$-axis). The position $\hat{\bm{r}}=(\sin\theta\cos\phi,
\sin\theta\sin\phi,\cos\theta)$ on the jetted region satisfies the
condition $\hat{\bm{r}} \cdot \bm{n}_{\rm j} \ge \cos\theta_{\rm j}$, where
$\bm{n}_{\rm j}=(\sin\theta_{\rm v}, 0, \cos\theta_{\rm v})$ is the direction
of the vector along the jet axis. This condition is rewritten as
%
\begin{eqnarray}
  \cos\phi
  \ge \frac{ \cos\theta_{\rm j}-\cos\theta_{\rm v}\cos\theta }{ \sin\theta_{\rm v}\sin\theta }
\equiv\cos\phi_{\rm ec},
\label{eq:visible}
\end{eqnarray}
%
and this leads to the egg missing part.
\par
The energy flux density of synchrotron emission should be considered based
on this egg. \cite{granot99} provided a general formula of the flux density
of radiation from a spherical expanding system for the case of an optically thin
limit and an isotropic radiation measured in the fluid frame,
%
\begin{eqnarray}
F(\nu,T) = && \frac{1+z}{4\pi d_{\rm L}{}^2}
\int_0^{2\pi} {\rm d}\phi \int_{-1}^{1} {\rm d}\mu \int_0^{\infty} r^2{\rm d}r \nonumber \\
&\times& \frac{ P'(\nu', \bm{r}, T_z+r\mu/c) }{ \gamma^2\left(1-\beta\mu\right)^2 },
\end{eqnarray}
%
where $d_{\rm L}$ is the luminosity distance to the GRB, $\beta=\sqrt{1-1/\gamma^2}$,
and $\nu'=\gamma\nu(1-\beta\mu)$ is the frequency of photon measured in the fluid
frame, respectively. The factor $1/\gamma^2\left(1-\beta\mu\right)^2$
represents the Doppler beaming effect, by which the bright region is concentrated to $\theta
\lesssim \gamma^{-1}$. The emission from the region at $\theta > \gamma^{-1}$ is beamed
away from the line of sight.
The emission power $P'$ depends on $\phi$ in our jet model:
$P'=0$ at which the inequality of Eq.~\eqref{eq:visible} is not satisfied.
Using the self-similar variables $\chi$ and $y$, we obtain
%
\begin{eqnarray}
F =  \frac{4R_l{}^3(1+z)}{\pi d_{\rm L}{}^2}
\int_0^{2\pi} {\rm d}\phi \int_{1}^{\chi_{\rm max}} {\rm d}
\chi \int_0^{\chi^{-\frac{1}{4}}} {\rm d}y
\frac{ \chi y^{10}P' }{ \left(1+7\chi y^4\right)^2 }. \nonumber \\
\end{eqnarray}
%
We take $\chi_{\rm max}=1+16\gamma_f{}^2$.
\par
The surface brightness profile (i.e. intensity image on the sky) at a given arrival
time $T$ is also derived. Let $R_{\perp}$ be a distance of a position on the sky
from the line of sight axis,
%
\begin{eqnarray}
R_\perp\equiv r\sin\theta
\simeq R_l y \sqrt{1-\mu^2}
\simeq \frac{\sqrt{2}R_l}{4\gamma_l}\sqrt{y-\chi y^5}.
\end{eqnarray}
%
The maximum value of the distance is given on the surface of the egg at which $(\chi,y)=
(1,5^{-\frac{1}{4}})$ as $R_{\perp,{\rm max}} \simeq 0.26 R_l/\gamma_l$. Its angular
size is $R_{\perp,{\rm max}}/(5^{-\frac{1}{4}}R_l) \sim 0.4\gamma_l^{-1}$.
Thus, the emission from the whole egg is bright due to the Doppler beaming.

Since the
entire size of the intensity image becomes larger with time due to the expansion of
the blast wave and its deceleration, it is convenient to introduce a variable of $x
\equiv R_{\perp}/R_{\perp,{\rm max}}$. The differential area of the image can be
written as
%
\begin{eqnarray}
{\rm d}S_{\perp} = R_\perp {\rm d}R_{\perp}{\rm d}\phi
= R_{\perp,{\rm max}}{}^2 x {\rm d}x{\rm d}\phi.
\end{eqnarray}
%
Thus, we obtain the flux density element in the image as
%
\begin{eqnarray}
\frac{ {\rm d}F }{ {\rm d}S_{\perp} }
&=&\frac{4}{\pi}\left(\frac{R_l}{d_{\rm L}}\right)^2
\frac{(1+z)^2}{cT}
\frac{ \chi y^{5}P' }{ \left(1+7\chi y^4\right)^2 } {\rm d}y \nonumber \\
&=&\frac{4}{\pi}\left(\frac{R_l}{d_{\rm L}}\right)^2
\frac{(1+z)^2}{cT}
\frac{ y^2\left(y-ax^2\right)P' }{ \left(8y-7ax^2\right)^2 } {\rm d}y,
\label{eq:flux density element}
\end{eqnarray}
%
where $a=8(\gamma_l R_{\perp,{\rm max}}/R_l)^2$ and we use $\chi=y^{-4}-ax^2 y^{-5}$. The
surface brightness profile is obtained after the $y$ integration as function of $R_\perp$
and $\phi$. The range of the $y$ integration is determined by the condition $\chi > 1$.
Note that the $\phi$ dependence of $P'$ (Equation~\ref{eq:visible})
can make the intensity image asymmetric (see Figure~\ref{fig:jet} and \ref{fig:schematic}).

\section{Synchrotron power and polarization}
\label{sec:synch}
\par
To calculate the synchrotron power at each point, we assume that the energies of electrons
$e_{\rm e}'$ and magnetic field $e_{\rm B}'$ are fixed fractions of the local internal
energy; $e_{\rm e}' = \epsilon_{\rm e}e'$ and $e_{\rm B}' = \epsilon_{\rm B}e'$. We suppose
that the electrons are accelerated by the forward shock and have a single power-law distribution
function everywhere:
%
\begin{eqnarray}
N(\gamma_{\rm e}) = K\gamma_{\rm e}{}^{-p}~~~{{\rm for}~\gamma_{\rm e}\ge\gamma_{\rm m}},
\end{eqnarray}
%
where $\gamma_{\rm e}$ is electron's Lorentz factor. The normalization constant $K$
and minimum Lorentz factor of electrons $\gamma_{\rm m}$ are determined as
%
\begin{eqnarray}
K = (p-1)n'\gamma_{\rm m}{}^{p-1},
\end{eqnarray}
%
and
%
\begin{eqnarray}
\gamma_{\rm m} = \left( \frac{p-2}{p-1} \right)
\frac{ \epsilon_{\rm e}e' }{ n' m_{\rm e}c^2 },
\end{eqnarray}
%
respectively. \cite{granot99} gives approximate synchrotron power formulas,
%
\begin{eqnarray}
P' =
\begin{cases}
P_{\nu,{\rm max}}'\left(\frac{\nu'}{\nu_{\rm m}'}\right)^{\frac{1}{3}} 
~~~{\rm for}~\nu' < \nu'_{\rm m} \\
P_{\nu,{\rm max}}'\left(\frac{\nu'}{\nu_{\rm m}'}\right)^{-\frac{p-1}{2}}
~~~{\rm for}~\nu' > \nu'_{\rm m}
\end{cases},
\label{eq:synch power}
\end{eqnarray}
%
where
%
\begin{eqnarray}
P_{\nu,{\rm max}}'=0.88\frac{ 4(p-1) }{ 3p-1 }
\frac{ n'P'_{\rm e,av} }{ \nu'_{\rm syn}\left(\langle\gamma_{\rm e}\rangle\right) }.
\end{eqnarray}
%
Here, $P'_{\rm e,av}$ is the synchrotron power by a single electron with an average Lorentz
factor of $\langle\gamma_{\rm e}\rangle \equiv \epsilon_{\rm e}e'/(n' m_{\rm e}c^2)$,
%
\begin{eqnarray}
P'_{\rm e,av}=\frac{4}{3}\sigma_{\rm T}c\beta_{\rm e}{}^2
\langle\gamma_{\rm e}\rangle^2 \epsilon_{\rm B}e',
\end{eqnarray}
%
where $\sigma_{\rm T}$ is the Thomson cross section, and $\beta_{\rm e}=\sqrt{1-1/
\langle\gamma_{\rm e}\rangle^{2}}$, respectively. The synchrotron peak frequency
measured in the fluid frame is
%
\begin{eqnarray}
\nu'_{\rm syn}(\gamma_{\rm e})=
\frac{ 3\gamma_{\rm e}{}^2 q_{\rm e} B' }{ 16m_{\rm e} c },
\end{eqnarray}
%
where $q_{\rm e}$ is the electron's electric charge and $B'=\sqrt{8\pi\epsilon_{\rm B}e'}$
is the local magnetic field strength. We define $\nu'_{\rm m}=\nu'_{\rm syn}(\gamma_{\rm m})$.
In this paper, we use this approximate power and fix the power-law index of electron's energy
spectrum to be $p=3$ for simplicity.
\par
%
\begin{figure}
\centering
\includegraphics[scale=0.3]{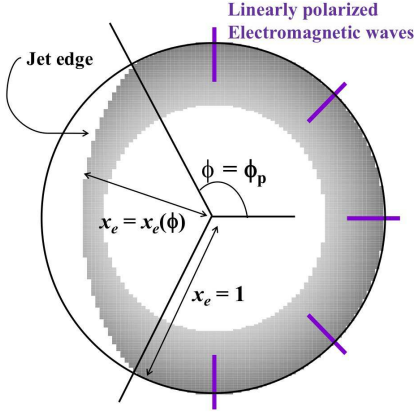}
\caption{Schematic picture of the asymmetry of the intensity image
and the distribution of local PAs. $\phi_{\rm p}$ denotes the azimuthal angle of
the region, which does not have a part missing due to the finite
extent of the jet. The angular distance from the line-of-sight axis to the outer edge
of nonzero intensity is denoted by $x_{\rm e}$.}
\label{fig:schematic}
\end{figure}
%
Synchrotron polarization depends on the magnetic field configuration at each position.
In this paper, we focus on the model in which the magnetic field is turbulent
with coherence length on the plasma scale, i.e., the power spectrum of magnetic turbulence ranges
only around the proton skin-depth scale, as done in \citet{sari99} and \citet{ghisellini99}.
Linear analysis and particle-in-cell simulations of relativistic collisionless shocks show that
the plasma-scale field perpendicular to the shock normal $\bm{B}'_\perp$ is predominantly generated
by Weibel instability at the shock \citep{medvedev99,gruzinov99}. The nonlinear evolution of
the field structure in the downstream of the shock is not fully understood, but the direction
of the distribution of the field may
be anisotropic \citep{nishikawa03,kato05,spitkovsky08,keshet09,sironi15}.
We parameterize its degree of anisotropy as
$\xi^2 \equiv 2\langle {B'_\parallel}^2 \rangle/\langle {B'_\perp}^2 \rangle$.
\par
Particle-in-cell simulations also show that nonthermal electron acceleration
is efficient in the plasma-scale field induced by Weibel instability when the unshocked medium
is not or weakly magnetized \citep{sironi11,sironi13}. For this reason we neglect
an ordered magnetic
field in our calculations. The strong helical magnetic field possibly inside the
jet cannot be
advected through the contact discontinuity and thus cannot affect the forward shock. 
\par
\begin{figure}
\centering
\includegraphics[scale=0.7]{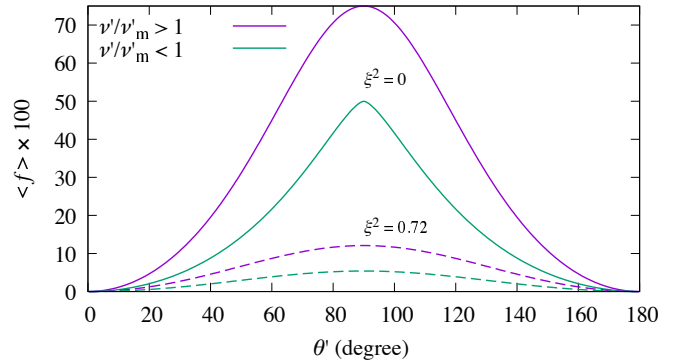}
\caption{The local PDs $\langle f\rangle$ as
functions of $\theta'$ for $\nu'/\nu_{\rm m}'>1$ (purple)
and $\nu'/\nu_{\rm m}'<1$ (green). The solid and dashed
lines represents $\xi^2=0$ and $\xi^2=0.72$, respectively.
Note that for $\gamma = 10$ as an example, the angular scales of
the Doppler-beamed region $\theta=1/\gamma$
and the whole emission region $\theta_{\rm j}+\theta_{\rm v}=9\;$degree
(as given in Section~\ref{sec:results}) correspond to $\theta'\simeq90\;$degree and $\simeq115\;$degree, respectively.}
\label{fig:fraction}
\end{figure}

Here, we derive the Stokes parameters of emission at each position, $\langle i' \rangle,
\langle q' \rangle, \langle u' \rangle$, averaged due to contributions from magnetic fields oriented
differently, by following \citet{sari99} \citep[see also][]{toma09,lan18}. Let us consider that a right-handed
coordinate system $x'y'z'$ whose axis $z'$ is along the photon wavevector $\bm{k}'$,
and the direction of
$\bm{B}'$ is described by spherical coordinates $(\theta'_{\rm B}, \phi'_{\rm B})$. Then the synchrotron
intensity depends on $\theta'_{\rm B}$ as $i' \propto \sin^\varepsilon \theta'_{\rm B}$, where
$\varepsilon = (p+1)/2$.
The direction of the polarization
is perpendicular to both $\bm{k}'$ and $\bm{B}'$, so that
\begin{equation}  
  q' = - i' f \cos 2\phi'_{\rm B}, ~~~
  u' = - i' f \sin 2\phi'_{\rm B},
\end{equation}
where $f=(p+1)/(p+7/3)$ \citep{rybicki79}. For another right-handed coordinate system of $1'2'3'$
whose axis of $3'$
is along the shock normal and
on the plane $x'z'$, we describe the direction of $\bm{B}'$ by spherical coordinates
$(\alpha',\eta')$. In the case of an anisotropic magnetic field
that has symmetry on the plane $1'2'$, we can
define a probability per unit of solid angle
for the field to have the polar angle $\alpha'$, $F_{\rm B}(\alpha')$, and
allow the field strength to depend on $\alpha'$, $B'=B'(\alpha')$ \citep{sari99,gill20}.
Recalling that the angle between axes $z'$ and $3'$
corresponds to the viewing angle $\theta'$, we have the relations 
%
\begin{eqnarray}
  \sin\theta_{\rm B}'\cos\phi_{\rm B}' &=& \sin\alpha' \cos\eta' \cos\theta' - \cos\alpha' \sin\theta' \nonumber \\
  \sin\theta_{\rm B}'\sin\phi_{\rm B}' &=& \sin\alpha' \sin\eta' \\
  \cos\theta_{\rm B}' &=& \sin\alpha' \cos\eta' \sin\theta' + \cos\alpha' \cos\theta' \nonumber
\end{eqnarray}
%
Averaging $i', q', u'$ with respect to the field distribution leads to $\langle u'
\rangle = 0$ due to the symmetry on the plane of $1'2'$ and gives the local PD as
\begin{eqnarray}
  \langle f \rangle &=& \frac{|\langle q' \rangle|}{\langle i' \rangle} \nonumber \\
  &=& \frac{|\int (-f \cos2\phi_{\rm B}')[B'(\alpha')\sin\theta_{\rm B}']^\varepsilon
             F_{\rm B}(\alpha') \sin\alpha' d\alpha' d\eta'|}
           {\int [B'(\alpha')\sin\theta_{\rm B}']^\varepsilon
             F_{\rm B}(\alpha') \sin\alpha' d\alpha' d\eta'}. \nonumber \\
\label{eq:fgeneral}
\end{eqnarray}
For the case of $p=3$ ($\nu'>\nu'_{\rm m}$), one has $\langle f \rangle$ without
specifying the functional forms of $F_{\rm B}(\alpha')$ and $B'(\alpha')$,
%
\begin{eqnarray}
\langle f \rangle = \frac{|\xi^2-1|\sin^2\theta'}{2+(\xi^2-1)\sin^2\theta'}.
\end{eqnarray}
%
For $\nu'<\nu'_{\rm m}$, we may adopt a model of the field distribution given by \cite{sari99},
%
\begin{eqnarray}
  B'(\alpha') \propto (\xi^2 \sin^2\alpha' + \cos^2\alpha')^{-1/2}, ~~~
  F_{\rm B}(\alpha') \propto {B'}^3(\alpha'). \nonumber \\
\end{eqnarray}
%
In this model, the distribution of $B'(\alpha')$ shapes an ellipsoid with aspect ratio $\xi$, and
$2 \langle {B'_\parallel}^2 \rangle/\langle {B'_\perp}^2 \rangle = \xi^2$ is consistently obtained
\citep{gill20}. For an extreme case of $\xi^2 = 0$, i.e., $\bm{B}' = \bm{B}'_\perp$, we have
$\langle f \rangle$ by setting $B'(\alpha') = $ const. and $F_{\rm B}(\alpha') =
\delta(\alpha' - \pi/2)$ in Equation~(\ref{eq:fgeneral}).
The calculation results of $\langle f \rangle$ for $\nu' > \nu'_{\rm m}$ and $\nu' < \nu'_{\rm m}$ are
shown in Figure~\ref{fig:fraction}. The solid and dashed lines represent $\xi^2 = 0$ and
$\xi^2 = 0.72$, respectively. We will show the calculation results
of polarization mainly for $\xi^2 = 0$, and
also show in Section~\ref{sec:nonzero_xi} that the typical PD measured in the late-phase
optical afterglows of $\sim 1-3\%$ is reproduced with
$\xi^2 = 0.72$ in the case of $\theta_{\rm v} = \theta_{\rm j}/2$.
\par
For $\xi^2 < 1$ we obtain $\langle q' \rangle > 0$.
This means that the direction of polarization is along axis
$x'$, which lies on the plane including $\bm{k}'$ and the shock normal. The Lorentz transformation of
radiation (i.e., the Doppler beaming) does not change this configuration, i.e., the direction
of the polarization
in the observer frame lies on the plane including $\bm{k}$ and the shock normal. As a result, the observed
polarization angle (PA) at the local point on the sky is identical to $\phi$ (see Figure~\ref{fig:schematic}).
%
%
\par
We obtain the surface brightness by the $y$ integration
obtain the surface brightness by the $y$ integration,
%
\begin{eqnarray}
S_I&=&
\int
\frac{ {\rm d}F }{ {\rm d}S_\perp {\rm d}y}
\langle \sin^{\varepsilon}\theta_{\rm B}' \rangle
{\rm d}y,
\\
S_Q&=&
\int
\langle f \rangle\frac{ {\rm d}F }{ {\rm d}S_\perp {\rm d}y}
\langle \sin^{\varepsilon}\theta_{\rm B}' \rangle
{\rm d}y
\cos2\phi,
\\
S_U&=&
\int
\langle f \rangle\frac{ {\rm d}F }{ {\rm d}S_\perp {\rm d}y}
\langle \sin^{\varepsilon}\theta_{\rm B}' \rangle
{\rm d}y
\sin2\phi.
\end{eqnarray}
%
\par
In summary, the synchrotron emission at each position is polarized due
to the anisotropy of turbulent magnetic field even though its coherence length is on
the plasma
skin-depth scale, many orders of magnitude smaller than the hydrodynamic scale, and the PA is
identical to $\phi$ (see Figure~\ref{fig:schematic}). The combination of the relativistic
aberration (or Doppler beaming) and off-axis viewing of the jet ($\theta_{\rm v} \neq 0$)
leads to an asymmetry in the observed image as explained in Section~\ref{sec:jet}. Therefore,
the emission integrated over the image can have net PD.

\section{Parameter setting}
\label{sec:parameter}
We perform calculations of PD temporal variations (which we call PD curves hereafter) and
PD spectra of GRB afterglows by using the above formulae with fixed parameters of the
isotropic energy of blast wave $E_{\rm iso} = 10^{52}\;$erg, $n_0 = 1\;{\rm cm}^{-3}$, $\epsilon_{\rm e}
= 0.1$, $\epsilon_{\rm B} = 0.01$, and $\theta_{\rm j} = 6\;$degree. They are typical
parameter values for long GRBs \citep[e.g.,][]{panaitescu02}. We assume nearby events
with $z \sim 0$.
The sideways expansion of collimated relativistic blast waves is
considered to be weak, as indicated by high-resolution hydrodynamic simulations \citep{zhang09,
vaneerten12}, so that we set $\theta_{\rm j} = {\rm const}.$, for simplicity. The Lorentz
factor of the blast wave is given by
%
\begin{eqnarray}
{\it \Gamma} = \frac{1}{2}\left(
\frac{17E_{\rm iso}}{16\pi n_0 m_{\rm p}c^5 T_z{}^3}\right)^{\frac{1}{8}}.
\end{eqnarray}
Then, for a given $\theta_{\rm v}$, we obtain observed synchrotron flux and polarization
as functions of $\nu$ and $T$.
\par
The magnetic field strength, proton plasma frequency, and electron cyclotron
frequency measured in the fluid frame are estimated as $B' = \sqrt{8\pi \epsilon_{\rm B} e'}\simeq
0.1 \;y^{-3/2}\chi^{-17/24}(T/1~{\rm day})^{-3/8}\;$G, $\omega_{\rm pp}' = \sqrt{4\pi q_{\rm p}{}^2
n'/\langle \gamma_{\rm p} \rangle m_{\rm p}}\simeq 3 \times 10^3 \chi^{-13/24}\;$Hz, and $\omega_{\rm ce}'
= q_{\rm e}B'/m_{\rm e}c \simeq 2 \times 10^6 y^{-3/2} \chi^{-17/24}(T/1~{\rm day})^{-3/8}\;$Hz, where
$q_{\rm p}$ is the proton's electric charge, and $\langle \gamma_{\rm p} \rangle = e'/n'm_{\rm p}c^2$
is the average Lorentz factor of thermal protons. The coherence length of turbulent magnetic field
induced by Weibel instability is on the plasma skin-depth scale, $c/\omega_{\rm pp}\sim10^7
\chi^{13/24}\;$cm, which is sufficiently small compared to the size of emission region, $R_{\perp,{\rm max}}
\simeq 0.26 R_l/\gamma_l \simeq 4 \times 10^{16} (T/1~{\rm day})^{5/8}\;$cm, that our treatment of averaging
$i', q', u'$ in Section~\ref{sec:synch} is justified.
%
\begin{figure}
\centering
\includegraphics[scale=0.3]{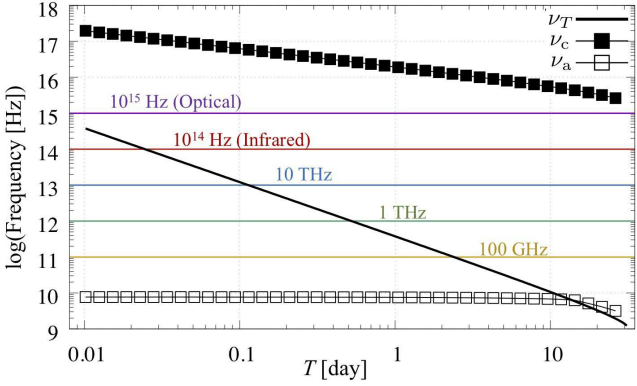}
\caption{Temporal variations of characteristic frequencies, $\nu_T=
\nu_{\rm m}(\chi=y=1)$ (black solid line), the synchrotron cooling
frequency $\nu_{\rm c}$ (filled squares), and the self-absorption
frequency $\nu_{\rm a}$ (open squares).}
\label{fig:frequencies}
\end{figure}
%
\par
The integrated flux spectrum has a broken power-law form because of
the $\nu$ dependence of synchrotron power (Equation~\ref{eq:synch power}) at each position, and
its peak frequency is given by
%
\begin{equation}
\nu_T = \nu_{\rm m}(\chi=y=1).   
\end{equation}
%
The image of the observed intensity tends to have a sharp ring shape for
$\nu/\nu_T>1$, while for $\nu/\nu_T<1$, the image tends to be a slab-like shape
\citep[cf.][and see also Figure~\ref{fig:images} below]{granot99}. Figure~\ref{fig:frequencies}
shows $\nu_T$ (black solid line) for our parameter choice during $T=0.01-30$~days.
$\nu_{\rm T}=\gamma_l(1+\beta)\nu_{\rm T}'$ evolves with time as
$\nu_T\propto \gamma_l \gamma_{\rm m}{}^2 B'\propto\gamma_l(\epsilon_{\rm e}e'/n')^2
\sqrt{\epsilon_{\rm B}e'}\propto\gamma_l^4\propto T^{-3/2}$.
\par
We consider the frequency range of $10^{11} - 10^{15}\;$Hz, at which the synchrotron cooling
and self-absorption are not important for our parameter set. Equating the synchrotron cooling
timescale of a single electron with the dynamical timescale $t'=R_l/(c\gamma_l)=16\gamma_l T$, we
obtain the cooling frequency measured in the observer frame at $\chi=y=1$ as
%
\begin{eqnarray}
\nu_{\rm c}= \gamma_l(1+\beta_l)\frac{3q_{\rm e}B'}{16m_{\rm e}c}
\left(\frac{6\pi m_{\rm e}c}{\sigma_{\rm T}B'^2 t'}\right)^2,
\end{eqnarray}
%
where $\beta_l=\sqrt{1-1/\gamma_l{}^2}$. In our setup, it remains at $\ga10^{15}$~Hz
during $T=0.01-30$~days, as displayed in Figure~\ref{fig:frequencies} (filled squares).
The synchrotron self-absorption coefficient at $\nu' < \nu'_{\rm m}$ can be estimated
as~\citep{rybicki79,granot99b}
%
\begin{eqnarray}
\alpha'_{\nu'}
=
\frac{3^{\frac{2}{3}} \sqrt{\pi}}{5\Gamma(\frac{5}{6})}
\frac{ (p+2)(p-1) }{ 3p+2 }
\frac{ q_{\rm e}{}^{\frac{8}{3}} }{ (m_{\rm e}c)^{\frac{5}{3}} }
n' B'^{ \frac{2}{3} }
\gamma_{\rm m}^{-\frac{5}{3}}
\nu'^{-\frac{5}{3}}~~~
\end{eqnarray}
%
where ${\rm \Gamma}(\zeta)$ is the gamma function. Then, we can estimate the absorption
frequency $\nu_{\rm a}$ as $0.247$ times the frequency at which $R_l\alpha'_{\nu'}/\gamma_l
= 1$, where $\alpha'_{\nu'}$ is taken at $\chi = y = 1$, and the numerical factor comes from
the effects of radiative transfer in the shocked fluid. We also plot $\nu_{\rm a}$ in
Figure~\ref{fig:frequencies} (open squares). It is constant for $\nu' < \nu'_{\rm m}$, while
$\nu_{\rm a} \propto T^{-(3p+2)/2(p+4)}$ for $\nu' > \nu'_{\rm m}$~\citep[e.g.][]{granot02}, and
as a result, it remains at $\la 10$~GHz. Hence, our calculations based on the synchrotron power
formula (Equation~\ref{eq:synch power}) are applicable from the submillimeter band (100~GHz) to
the optical band ($10^{15}$~Hz).

\section{Results}
\label{sec:results}
%
\begin{figure*}[htbp]
\centering
\includegraphics[scale=1.3]{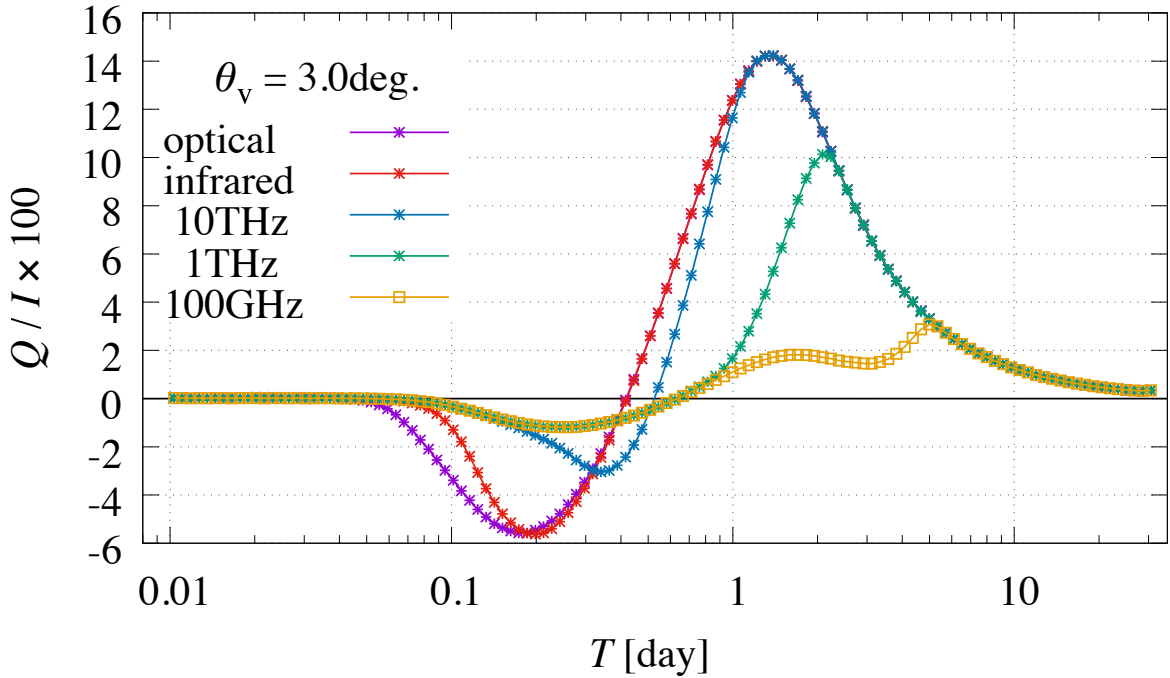}
\caption{PD curves for $\xi^2 = 0$ and $\theta_{\rm v}=3$~degree, at frequencies
$10^{15}$~Hz (optical), $10^{14}$~Hz (infrared), 10~THz, 1~THz, and
100~GHz during $T=0.01$-$30$ days. The 100~GHz PD at the initial peak
is smaller than that of optical band by a factor of $\sim 5$, but
begins to approach the optical curve once $\nu/\nu_T\ga1$. The two curves
overlap when $\tilde{S}_Q/\tilde{S}_I$ becomes identical at the two
bands (see Figure~\ref{fig:contrast}).}
\label{fig:polcurve}
\end{figure*}
%
We show the calculation results of synchrotron polarization from collimated blast waves as
functions of $T$ and $\nu$ mainly for the case of $\xi^2 = 0$ and their
interpretation in this section. Figure~\ref{fig:polcurve}
represents the calculated PDs,
%
\begin{eqnarray}
\frac{Q}{I} = \frac{
\int_{-\pi}^{\pi} {\rm d}\phi\int_0^{1} S_Q x{\rm d}x}
{\int_{-\pi}^{\pi} {\rm d}\phi\int_0^{1} S_I x{\rm d}x}
\label{eq:Q/I}
\end{eqnarray}
%
for the given $\theta_{\rm v}=3$~degree, at frequencies of
$10^{15}$~Hz (optical), $10^{14}$~Hz (infrared),
10~THz, 1~THz, and 100~GHz during $T=0.01-30$ days (note that $U=0$). We find that the temporal behaviors
of PDs in the optical and radio bands are basically the same in the sense that the PD curves have two
peaks, but the radio PD is significantly lower than the optical one with their ratio at the first PD
peaks is $\sim 5$. Moreover, the PDs of radio bands begin to approach that of optical band once $\nu
/\nu_T \gtrsim 1$, at $T \sim 0.5\;$day for $1\;$THz and $T \sim 3\;$day for $100\;$GHz (see
Figure~\ref{fig:frequencies}).
%
\begin{figure*}[htbp]
\centering
\includegraphics[scale=0.5]{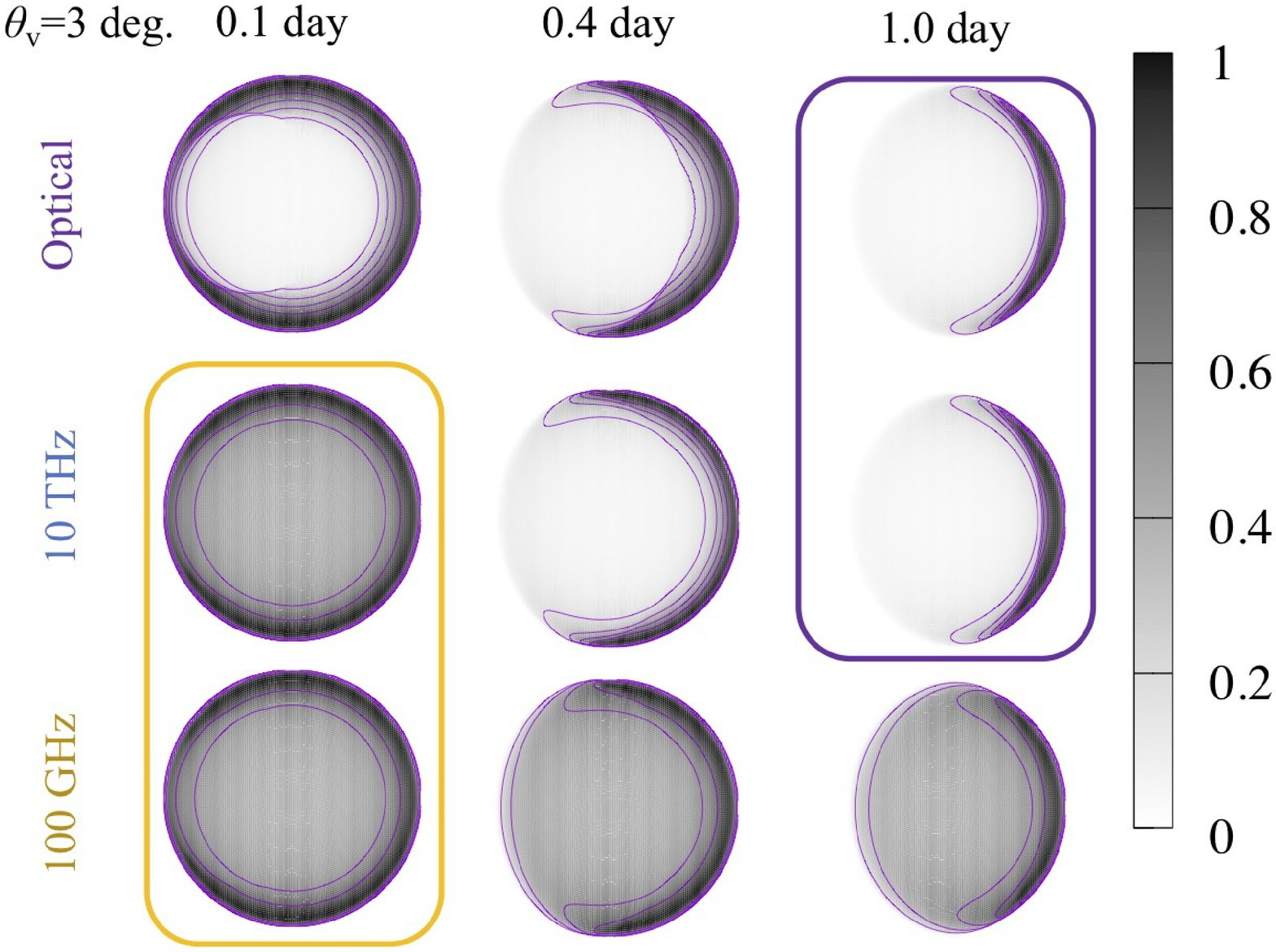}
\includegraphics[scale=0.5]{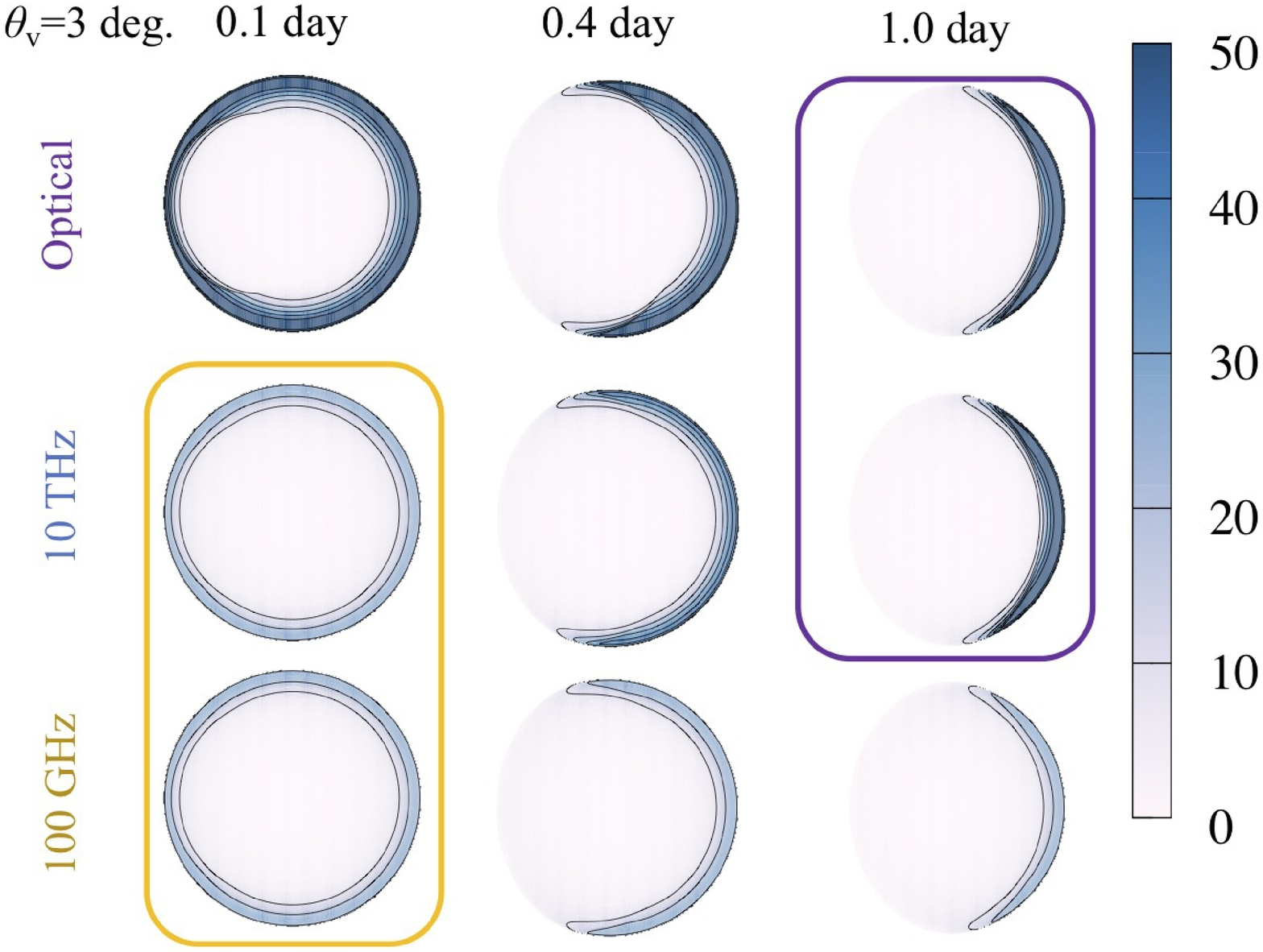}
\caption{Surface brightness of $S_I$ (top) and $\sqrt{{S_Q}^2+{S_U}^2}$
(bottom) for $\nu=10^{15}$~Hz (optical), 10~THz and 100~GHz with $\xi^2 = 0$ and $\theta_{\rm v}
=3$~degree at $T= 0.1, 0.4,$ and $1.0$~day. The Stokes $S_I$ is normalized by
its maximum value $S_{I, {\rm max}}$, while as for the Stokes $\sqrt{{S_Q}^2
+{S_U}^2}$, we display $\sqrt{{S_Q}^2+{S_U}^2}/S_{I, {\rm max}}\times100$.
The image at 10~THz is almost the same as the 100~GHz image ($\nu_{100{\rm GHz}}
/\nu_T\ll1$) at $T=0.1~$day, approaches the optical image ($\nu_{\rm opt}/
\nu_T \gg 1$) at $T=0.4~$day, and becomes almost the same as the optical
image at $T= 1.0~$day.}
\label{fig:images}
\end{figure*}
%
\par
Figure~\ref{fig:images} shows the images of surface brightness $S_I$ at the frequencies of $10^{15}$~Hz
(optical; $\nu/\nu_T \gg 1$ case), $10$~THz, and $100$~GHz ($\nu/\nu_T \ll 1$ case) for the
given observed
times $T= 0.1, 0.4,$ and $1.0\;$ day and viewing angle
$\theta_{\rm v}=3$~degree. A missing part gets larger
with time in each wave band due to the finite extent of the jet and the growth of the egg size.
The image shape at $10\;$THz is similar to that at $100\;$GHz
at $T = 0.1\;$day, but becomes similar to that at
the optical band at $T = 1\;$day. This behavior corresponds to that of the PD curves shown in Figure~\ref{fig:polcurve}.
\par
We note that even if we consider a nearby event, say at a distance $\sim 100$~Mpc,
the angular size of its image is $\sim R_{\perp,{\rm max}}/(100~{\rm Mpc})\sim 30 (T/1{\rm day})^{5/8}\; \mu$as.
The imaging polarimetry of such a transient object is difficult for the current telescopes. Thus we calculate
the PD curves and PD spectra by integrating the polarized flux over the image (Equation~\ref{eq:Q/I}), while
we show the (polarized) intensity images (Figure~\ref{fig:images}) just for explaining physical mechanisms.
Below we will explain the details of relationship between the PD curves and the asymmetry of images.

\subsection{Temporal behaviors of PDs}
\label{sec:behaviour}
%
\begin{figure*}[t]
\centering
\includegraphics[scale=0.6]{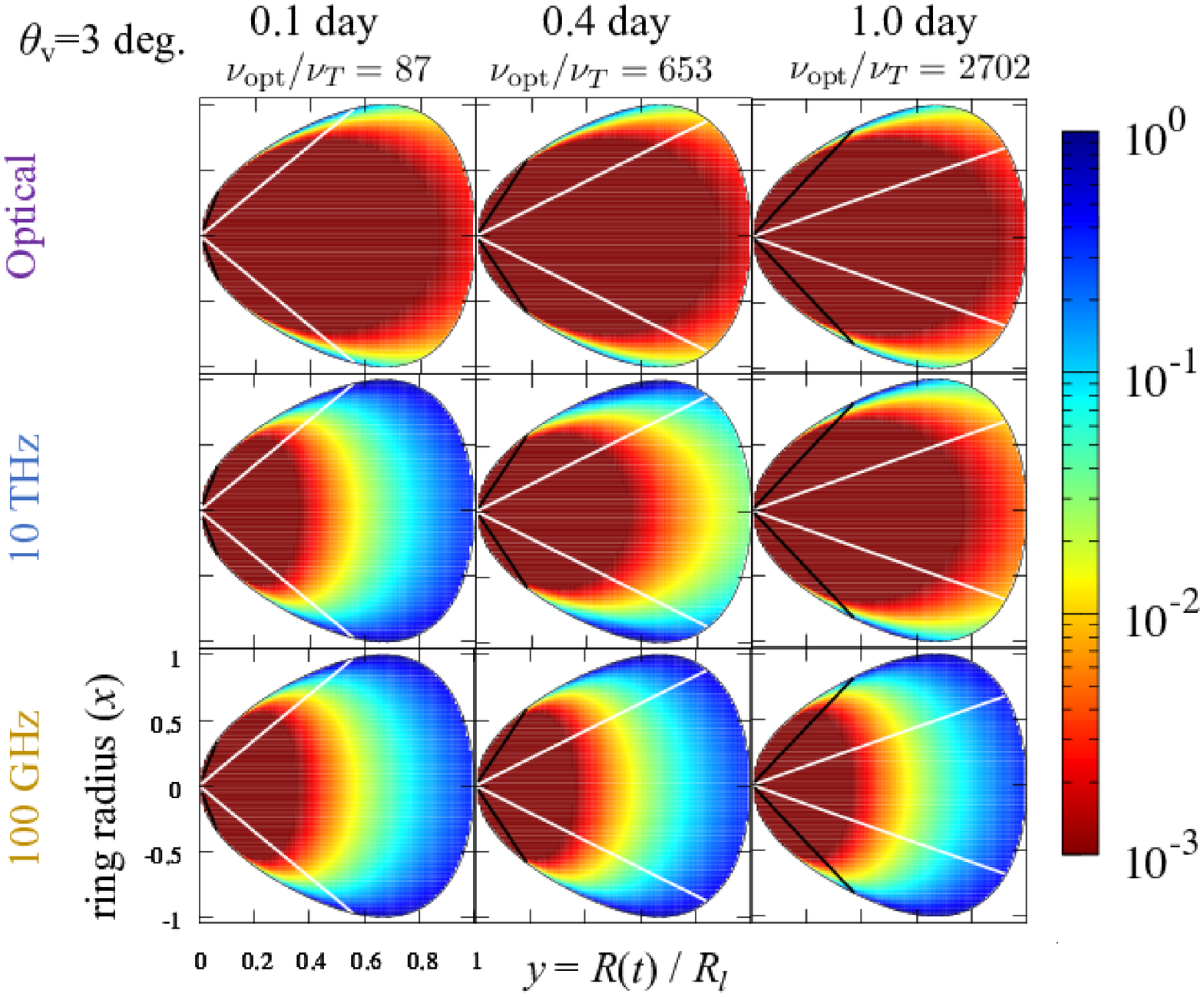}
\caption{The flux density element ${\rm d}F/{\rm d}S_{\perp} {\rm d}y$ (in
arbitrary units) in the $x$-$y$ plane for the same parameter values as for
Figure~\ref{fig:images}. The black and white lines are $\cos\phi_{\rm ec}
(x,y)=1$ and $\cos\phi_{\rm ec}(x,y)=-1$ (white line), respectively. The
region between the two white lines corresponds to $\cos\phi_{\rm ec}<-1$
($P'\ne0$ for an arbitrary $\phi$), between the white and black lines
corresponds to $-1<\cos\phi_{\rm ec}<1$ ($P'$ depends on $\phi$), and the
other region corresponds to $\cos\phi_{\rm ec}>1$ ($P'=0$ for an arbitrary
$\phi$).}
\label{fig:egg}
\end{figure*}
%
Firstly, we review the temporal behavior of optical PD~\citep[][]{sari99,ghisellini99,lazzati06}:
The PD curve has two peaks, and the PA rotates by 90 degree between the two peaks. This can be
understood by the temporal change of the image shown in Figure~\ref{fig:images} (top).
The shape of the optical
image is always like a ring (or an
asymmetric ring) because of $\nu/\nu_{\rm T} \gg 1$
\citep{granot99}. Figure~\ref{fig:egg} represents the flux density element ${\rm d}F/{\rm d}
S_\perp{\rm d}y$, which shows that the bright ring is produced by the blue region in the egg.
The lines of $\cos\phi_{\rm ec}(x,y)=1$ (black line) and $\cos\phi_{\rm ec}(x,y)=-1$ (white line)
are also shown in the $x$-$y$ plane. The region between the two white lines corresponds to $\cos
\phi_{\rm ec}<-1$ ($P'\ne0$ for an arbitrary $\phi$), that between the white and black lines to
$-1<\cos\phi_{\rm ec}<1$ ($P'$ depends on $\phi$), and the other region to $\cos\phi_{\rm ec}>1$
($P'=0$ for an arbitrary $\phi$). From these figures we can see that at $T \lesssim 0.1\;$ day
the optical image does not show a clear asymmetry. Figure~\ref{fig:images} (bottom) shows that the
surface polarized brightness $\sqrt{{S_Q}^2+{S_U}^2}$ is also large on the ring. The local PA is
identical to $\phi$, as illustrated in Figure~\ref{fig:schematic}. Therefore, since we observe the
emission from a GRB as a point source, the net PD is negligible due to the axisymmetry of the local
PA distribution. Then, as time goes, the optical image is
asymmetric (i.e., having a missing part), and the symmetry breaking gives
rise to nonzero PD.
By considering the superposition of linearly polarized waves, we can see that the observed PD has
one local maximum at $\phi_{\rm p} \sim 3\pi/4$ (the first peak) and zero PD at $\phi_{\rm p} \sim
\pi/2$ (rotation of PA). The PD decays after the second maximum when most of the bright ring is
missing (see more quantitative explanation in Section~\ref{sec:quantitative}).
\par
The PD curve at 100~GHz is similar to the optical PD curve until $T \sim 3$~day at which $\nu/\nu_T
\sim1$ (see Figure~\ref{fig:frequencies}), but the peak PD is lower. Figure~\ref{fig:images} (bottom)
shows that the 100\;GHz image is not ring-like because of $\nu/\nu_{\rm T} \ll 1$, in contrast to
the optical image. From Figure~\ref{fig:egg} we can see that a larger area with $0 < x <1$ contributes
to the observed intensity image compared to the optical case~\citep{granot99}. In this central part
the surface polarized brightness $\sqrt{{S_Q}^2+{S_U}^2}$ is small as can be
seen in Figure~\ref{fig:images}
(bottom). Therefore, the total emission at $100\;$GHz has a net PD lower than that at the optical band.
\par
The 100~GHz PD curve begins to approach the optical curve once $\nu/\nu_T\ga1$, synchronizing with the
behavior of the intensity images, i.e., the image shape becomes similar to that of optical. This also
explains the behaviors of PD curves at the other bands (1~THz, 10~THz, and infrared). Interestingly,
when $\nu/\nu_T$ becomes unity after the second peak, the curves show the third peaks as a result of
the approach to the optical curve.

\subsection{Further Quantitative Understanding}
\label{sec:quantitative}
To understand the PD curves more quantitatively, we derive analytical estimate of the PDs. First of all,
as shown in Figure~\ref{fig:schematic}, we regard $\phi_{\rm p}$ as a representative azimuthal angle
dividing the intensity image into the region having a missing part ($-\pi\le\phi\le-\phi_{\rm p}$ and
$\phi_{\rm p} \le \phi\le\pi$) and the rest ($-\phi_{\rm p}\le\phi\le\phi_{\rm p}$).
Here, we set $\phi_{\rm p}
\sim\phi_{\rm ec}(y_{\rm e})$, where $y_{\rm e}=5^{-\frac{1}{4}}$ denotes the edge of
the egg $(x,y)=(1,y_{\rm e})$
($(\chi,y)=(1,y_{\rm e})$), for simplicity, although the brightest part on $\chi=1$ is slightly different.
We define $x_{\rm e}(\phi)$ as the angular distance of the outer edge of bright part from the line of sight.
In the region without the missing part, $x_{\rm e}$ is always unity, while in the
region with the missing part, $x_{\rm e}$ is a function of $\phi$ in general. Next, for the
region without the missing part we define $\tilde{S}_I \equiv \int_0^1 S_I x{\rm d}x$
and $\tilde{S}_Q \equiv \int_0^1 (S_Q/\cos2\phi) x{\rm d}x$, and
approximate $\tilde{S}_I$ and $\tilde{S}_Q$
as $\phi$ independent by ignoring the detailed structure
of the images (e.g. the optical image at $T = 0.4\;$day
shown in Figure~\ref{fig:images}). This approximation leads to a rough estimate of $Q/I$ by which we can catch
some essential points on the multi-wave band polarization. Then we obtain
%
\begin{eqnarray}
I&=&\int_{-\pi}^{\pi}{\rm d}\phi\int_0^1 S_I x{\rm d}x \nonumber \\
&\sim& \int_{-\phi_{\rm p}}^{\phi_{\rm p}}{\rm d}\phi\int_0^1 S_I x{\rm d}x
+2\int_{\phi_{\rm p}}^{\pi}{\rm d}\phi\int_0^{x_{\rm e}(\phi)} S_I x{\rm d}x,
\nonumber
 \\
&\sim& 2\phi_{\rm p}\tilde{S}_I
+2\int_{\phi_{\rm p}}^{\pi}\Delta\tilde{S}_I{\rm d}\phi,
\end{eqnarray}
%
%
\begin{eqnarray}
Q&=&\int_{-\pi}^{\pi}{\rm d}\phi\int_0^1 S_Q x{\rm d}x \nonumber \\
&\sim& \int_{-\phi_{\rm p}}^{\phi_{\rm p}}{\rm d}\phi\int_0^1 S_Q x{\rm d}x
+2\int_{\phi_{\rm p}}^{\pi}{\rm d}\phi\int_0^{x_{\rm e}(\phi)} S_Q x{\rm d}x,
\nonumber
\\
&\sim&\tilde{S}_Q\sin{2\phi_{\rm p}}
+2\int_{\phi_{\rm p}}^{\pi}\Delta\tilde{S}_Q\cos{2\phi}{\rm d}\phi,
\end{eqnarray}
%
and $U=0$.
Here, we have also defined $\Delta\tilde{S}_I \equiv \int_0^{x_{\rm e}(\phi)} S_I x{\rm d}x$ and
$\Delta\tilde{S}_Q \equiv \int_0^{x_{\rm e}(\phi)} (S_Q/\cos2\phi) x{\rm d}x$ for the
region with the missing part.
Let us further approximate $\Delta\tilde{S}_I$ and $\Delta\tilde{S}_Q$ as $\phi$
independent, the observed PD is finally written as
%
\begin{eqnarray}
\frac{Q}{I}\sim \frac{\tilde{S}_Q}{\tilde{S}_I}
\frac{\sin{2\phi_{\rm p}}}{2\phi_{\rm p}}
\frac{1-\Delta\tilde{S}_Q/\tilde{S}_Q}
{1+\left(\pi/\phi_{\rm p}-1\right)
\Delta\tilde{S}_I/\tilde{S}_I}.
\label{eq:approximate}
\end{eqnarray}
%
Since $\Delta\tilde{S}_Q/\tilde{S}_Q\ll1$ and $\Delta\tilde{S}_I/\tilde{S}_I\lesssim1$ (see
Figure~\ref{fig:images}), the rightmost factor is around unity. The factor $\tilde{S}_Q/
\tilde{S}_I$ represents the wavelength dependence of $Q/I$ through the wavelength dependences
of the $S_I$ and $\sqrt{{S_Q}^2+{S_U}^2}$ images shown in Figure~\ref{fig:images}. The factor
$\sin{2\phi_{\rm p}}/2\phi_{\rm p}$ describes the cancellation of superposition of local linear
polarizations for our present case of radial local PAs, and leads to the first peak PD at
$\phi_{\rm p} \sim 3\pi/4$ and the rotation of PA at $\phi_{\rm p} \sim \pi/2$.
%
%
\begin{figure}
\includegraphics[scale=0.3]{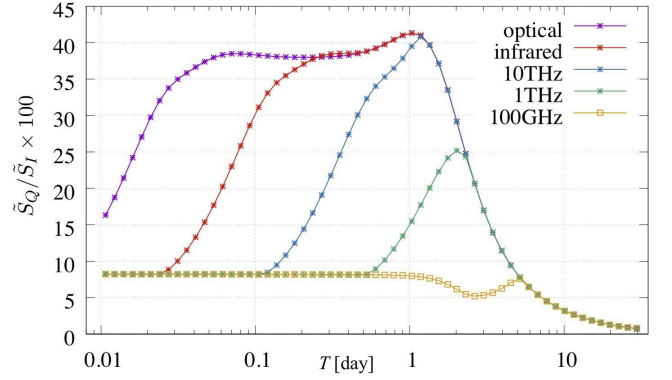}
\caption{Temporal variation of the factor
$\tilde{S}_Q/\tilde{S}_I$ for $\phi = 0$ at
the multiple wave bands for the same parameter
values as Figure~\ref{fig:polcurve}.}
\label{fig:contrast}
\end{figure}
\par
Figure~\ref{fig:contrast} shows the calculation result of $\tilde{S}_Q/\tilde{S}_I$ for
$\phi = 0$ at the multiple wave bands for the same parameter values as Figure~\ref{fig:polcurve}.
At the optical band, $\tilde{S}_Q/\tilde{S}_I \sim {\rm const}.$ for $0.1 \lesssim T \lesssim 1\;$day.
The decay time of $\tilde{S}_Q/\tilde{S}_I$ ($T \sim 1\;$ day at optical) produces the second PD peak.
This is the time at which $\theta_{\rm j} + \theta_{\rm v} \sim 0.6 \gamma_l^{-1}$, which corresponds
to the so-called jet break time $T_{\rm j}$ \citep[e.g.][]{sari99,ghisellini99}, when most of the
bright highly polarized ring emission is missing. The factor $\sin2\phi_{\rm p}/2\phi_{\rm p}$ explains
the behavior of the PD curve; the ratio of
$\sin2\phi_{\rm p}/2\phi_{\rm p} \sim -0.2$ at $\phi_{\rm p} \sim
3\pi/4$ to $\sim 0.4$ at $\phi_{\rm p} \sim \pi/3$ (at $T = T_{\rm j}$) is consistent with the ratio of
the first peak PD, $Q/I \simeq -6\%$, to the second peak PD, $Q/I \sim 14\%$ (Figure~\ref{fig:polcurve}).
At the other wave bands, $\tilde{S}_Q/\tilde{S}_I$ increases once $\nu/\nu_T\ga1$, from which the PD
curves begin to approach the optical PD curve, as shown in Figure~\ref{fig:polcurve}. At the time
$\tilde{S}_Q/\tilde{S}_I$ overlaps with that of optical band, the PD curves also overlap. We can
understand the difference in PD between the optical band and 100~GHz band by the difference in
$\tilde{S}_Q/\tilde{S}_I$; this factor is different by a factor of $\sim5$ around the PD peak times, which
agrees with the numerical results of PDs (e.g. the first peak PDs $Q/I \simeq -6\%$ at the optical band
and $Q/I \simeq -1\%$ at 100~GHz).

\subsection{PD curves for different $\theta_{\rm v}$}
\label{sec:PD curves view}
\par
%
\begin{figure}
\centering
\includegraphics[scale=0.7]{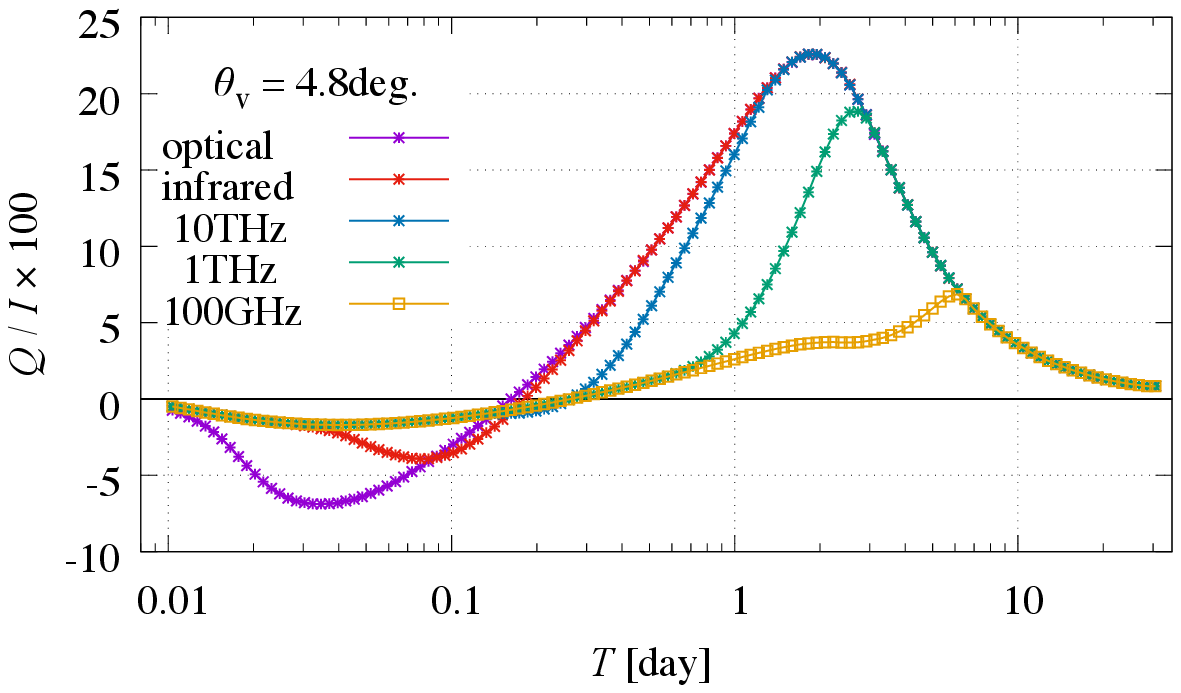}
\includegraphics[scale=0.7]{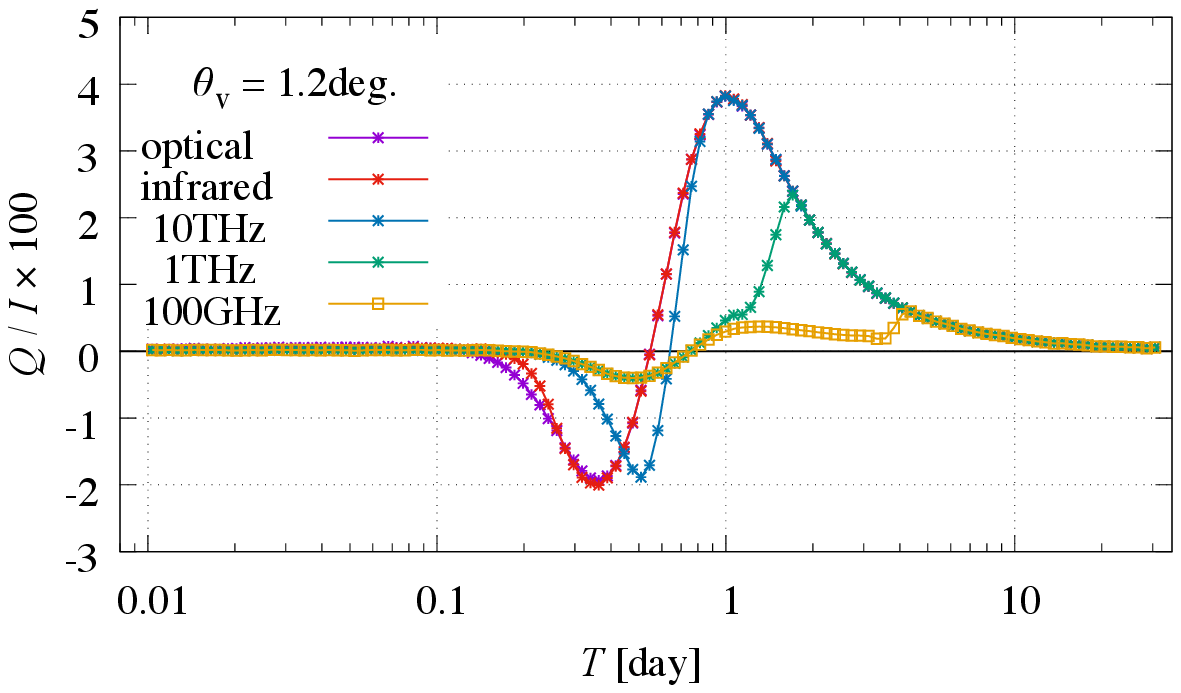}
\caption{PD curves for $\theta_{\rm v} = 4.8\;$degree
(top panel) and 1.2\;degree (bottom panel) with the same values as
the other parameters as Figure~\ref{fig:polcurve}.}
\label{fig:pdcurve48}
\end{figure}
%
Figure~\ref{fig:pdcurve48} (top) shows
the multi-wave band PD curves calculated for $\theta_{\rm v} =0.8\;$degree
and $\theta_{\rm j}=4.8\;$degree. The overall behaviors are similar to those for $\theta_{\rm v} =
3\;$degree (Figure~\ref{fig:polcurve}), but the image begins to be asymmetric earlier, and the PA flips ($\phi_{\rm p}
\sim \pi/2$) also occur earlier than the case of $\theta_{\rm v} = 3\;$degree. On the other hand, the jet
break time (when $\theta_{\rm j} + \theta_{\rm v} \sim 0.6\gamma_l^{-1}$) is later. The value of $\tilde{S}_Q
/\tilde{S}_I$ is the same as that for $\theta_{\rm v} = 3\;$degree, but its decay time is $T \simeq 2\;$day. Thus,
the PD is determined mainly by $\sin2\phi_{\rm p}/2\phi_{\rm p}$, and the later jet break time (at which
$\sin2\phi_{\rm p}/2\phi_{\rm p} \sim 0.6$) makes the second peak PDs higher than those for $\theta_{\rm v}
= 3\;$degree. The ratio of the first peak PDs between the optical and 100~GHz is $\sim 5$ also in this
case due to the same value of $\tilde{S}_Q/\tilde{S}_I$.
\par
Figure~\ref{fig:pdcurve48} (bottom panel) shows
the multi-waveband PD curves calculated for $\theta_{\rm v} =
0.2 \theta_{\rm j}=1.2\;$degree. In this case the image begins
to be asymmetric later and the jet break time is earlier
than in the case of $\theta_{\rm v} = 3\;$degree.
The value of $\tilde{S}_Q/\tilde{S}_I$ also does not change also
in this case, but its decay time is $T \simeq 0.6\;$day. This makes the second
peak PDs lower than those
for $\theta_{\rm v} = 3\;$degree. We should note that
for the optical and infrared bands, the first PD peak
time of $T \sim 0.4\;$day is significantly earlier than the peak time of $\sin2\phi_{\rm p}/2\phi_{\rm p}$
($T \sim 0.6\;$day). This is mainly due to our approximation $\phi_{\rm p} \sim \phi_{\rm ec}(y_{\rm e})$
for the rough estimate of $Q/I$. The brightest part of ${\rm d}S_\perp/{\rm d}\chi {\rm d}y$ is at $y <
y_{\rm e}$ (see Figure~\ref{fig:egg}).

\subsection{PD Spectra}
%
\begin{figure}
\centering
\includegraphics[scale=0.7]{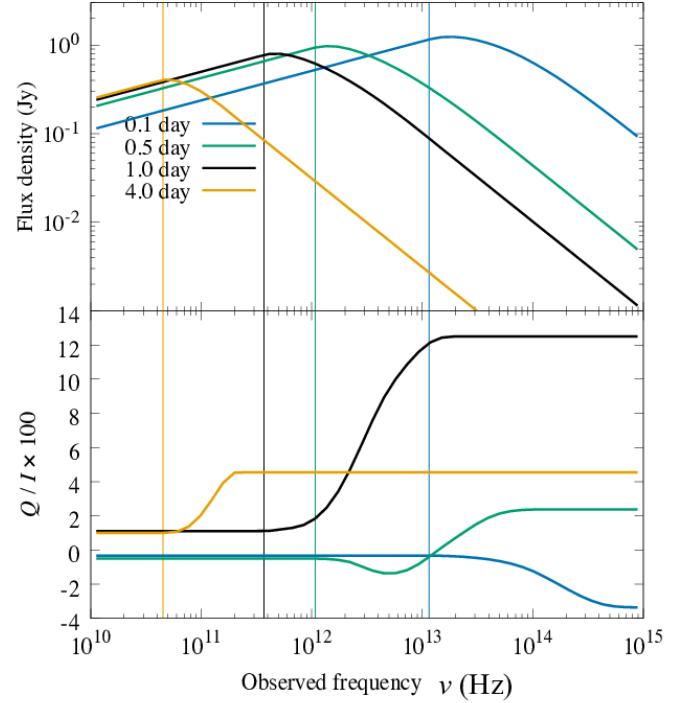}
\caption{The top panel shows the spectra of Stokes~$I$ for $\xi^2 = 0$ and
$\theta_{\rm v}=3$~degree at $T=0.1$ (blue), 0.5 (green),
1.0 (black), and 4.0 (yellow)~days. We assume a nearby GRB
with a luminosity distance of $d_{\rm L}=100$~Mpc. The bottom
panel shows the PDs. The vertical thin lines indicate $\nu
/\nu_T=1$ at each time.}
\label{fig:pinu}
\end{figure}
%
Figure~\ref{fig:pinu} shows spectra of
the observed Stokes~$I$ and PDs for $\xi^2 = 0$ and $\theta_{\rm v}=3$~degree
at $T=0.1$, 0.5, 1.0, and 4.0~days. Here, we consider a nearby GRB event with
a luminosity distance of
$d_{\rm L}=100$~Mpc. The PDs at $\nu < \nu_T$ are always lower than that at optical bands ($\nu \gg
\nu_T$) because of the difference in $\tilde{S}_Q/\tilde{S}_I$. The PA at 100~GHz is the same as
that at the optical band at most of the times as shown by the PD spectra at $T = 0.1$ and $1.0$
days, but they can be different by 90~degrees as shown by the PD spectrum at $T = 0.5\;$day within
a short time period. This period is $T \sim 0.4 - 0.7\;$day as indicated by the PD curves (Figure~\ref{fig:polcurve}).
These are characteristic properties of the plasma-scale turbulent magnetic field model, so that
simultaneous polarimetric observations at the optical and radio bands would provide a firm test
of this model. In other words, if the PDs at the radio bands of $\nu < \nu_T$ are higher than
that at the optical band or the difference in PAs at the two bands is not 0 or 90 degree, we can
rule out the plasma-scale turbulent field model.
\par
The PD spectra are flat at $\nu < \nu_T$ (low PD regime) and $\nu > \nu_B$ (high PD regime), while
the PD gradually varies at $\nu_T < \nu < \nu_B$, with $\nu_B \sim 40 \nu_T$ at $T < T_{\rm j}$.
This behavior can be explained by the variation of $\tilde{S}_I/\tilde{S}_Q$ in terms of $\nu$ with
fixed $T$ (see Figure~\ref{fig:contrast}). At $T > T_{\rm j}$, $\nu_B/\nu_T$ rapidly decreases (e.g.
$\nu_B/\nu_T \sim 5$ at $T = 4.0\;$day). The ratio $\nu_B/\nu_T$ at $T < T_{\rm j}$ does not significantly
depend on $\theta_{\rm v}$ because of the same value of $\tilde{S}_Q/\tilde{S}_I$. We should note, however,
that the detailed shapes of PD spectra are not explained only by
the wavelength dependence of $\tilde{S}_Q
/\tilde{S}_I$. For example, at $T = 0.1\;$day the PD at infrared band ($10^{14}$~Hz) is much lower
than that of the optical band, but the $\tilde{S}_Q/\tilde{S}_I$ of infrared band is already comparable
to that of the optical band. In such cases, the differences in PDs are affected by the third factor in
Eq.~(\ref{eq:approximate}) or more accurately by the combination of the flux density element
${\rm d}F_\nu/{\rm d}S_{\perp}{\rm d}y$ profile and $\cos\phi_{\rm ec}$.
%
\begin{figure}
\centering
\includegraphics[scale=0.7]{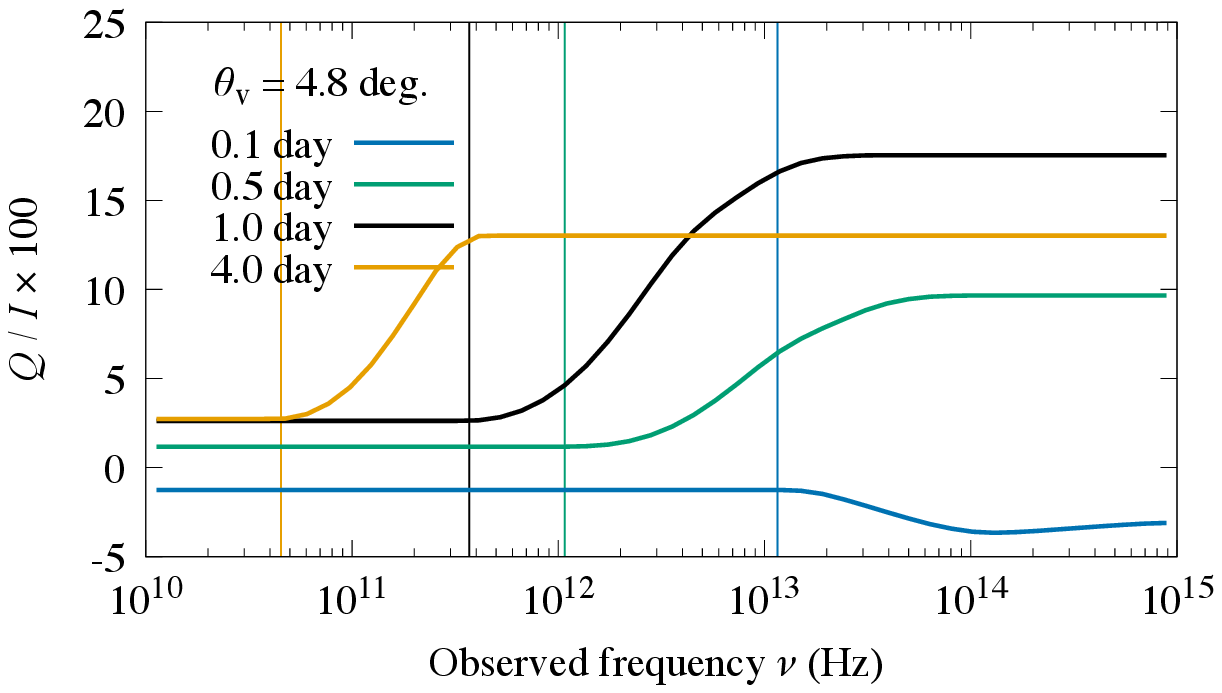}
\includegraphics[scale=0.7]{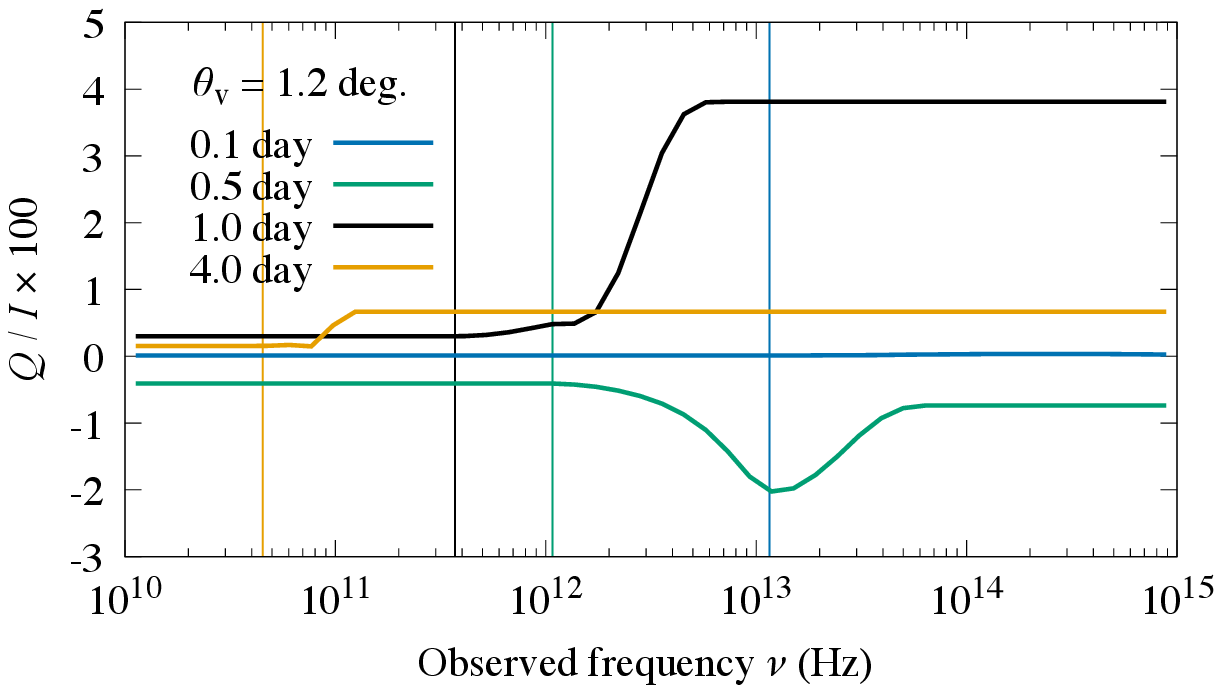}
\caption{The PD spectra for $\theta_{\rm v}=4.8$~degree
(top panel) and 1.2~degree (bottom) panel. The vertical thin lines indicate $\nu
/\nu_T=1$ at $T=0.1$ (blue), 0.5 (green), 1.0 (black), and 4.0 (yellow)~days.}
\label{fig:pinu view}
\end{figure}
%
\par
Figure~\ref{fig:pinu view} represents the PD spectra for $\theta_{\rm v}=4.8\;$degree
(top panel)
and $\theta_{\rm v}=1.2\;$degree (bottom panel).
The spectra are also flat at $\nu<\nu_T$ and $\nu>\nu_B$.
At $\nu_T<\nu<\nu_B$, the curves are different from the case of $\theta_{\rm v}=3.0\;$degree.
This is due to the difference in times at which
the images begin to be asymmetric (see Section~\ref{sec:PD curves view}).

\subsection{PD curves and spectra for $\xi^2=0.72$}
\label{sec:nonzero_xi}
%
\begin{figure}
\centering
\includegraphics[scale=0.7]{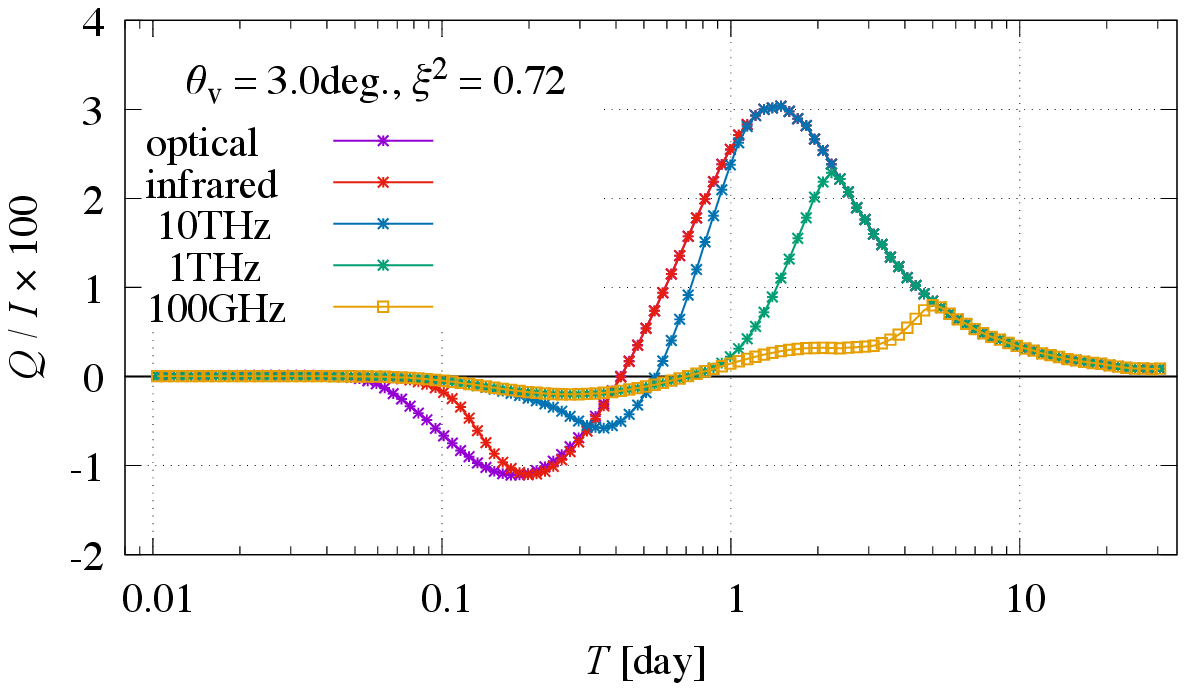}
\includegraphics[scale=0.7]{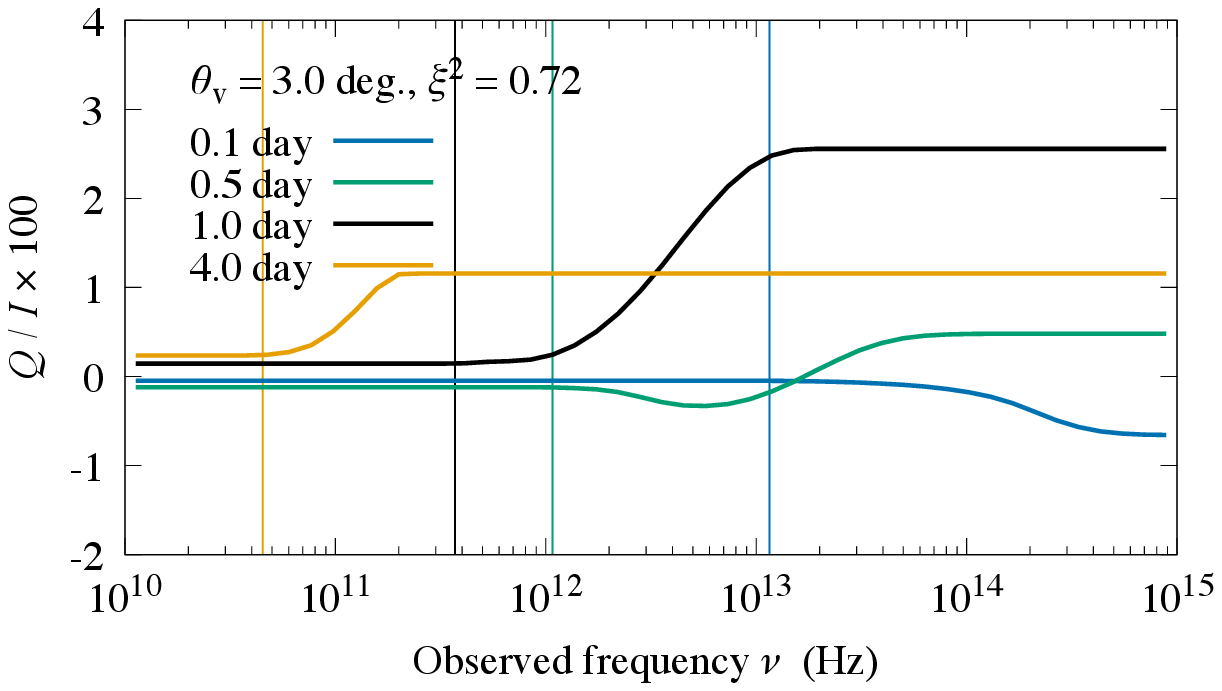}
\caption{The top panel is the PD curves for $\xi^2=0.72$ and $\theta_{\rm v}=3\;$degree.
The bottom panel is the PD spectra at $T=0.1$ (blue), 0.5 (green),
1.0 (black), and 4.0 (yellow)~days. The vertical thin lines indicate $\nu
/\nu_T=1$ at each time.}
\label{fig:pol xi}
\end{figure}
%
We also calculate the PD curves and spectra for
$\xi^2=2\langle B_\parallel '^2\rangle/\langle B_\perp'^2\rangle=0.72$
and $\theta_{\rm v}=\theta_{\rm j}/2=3\;$degree. The results are shown in Figure~\ref{fig:pol xi}.
The less anisotropic magnetic field structure leads to a
smaller $\langle f \rangle$ (see Figure~\ref{fig:fraction})
and weaker net PDs. The value of $\xi^2 = 0.72$ reproduces the observed typical late-phase optical
PD $\sim 1-3\%$. The ratio of PD peak values in the optical and radio bands is $\sim 10$. The temporal behaviors
of multi-band intensity image shapes are almost the same as the case of $\xi^2 = 0$, which provide almost
the same times of PD peaks and PA $90^\circ$ changes and values of $\nu_B$.

\section{Summary and Discussion}
\label{sec:discussion}
We have studied multi-wave band
polarization of GRB afterglows under the assumption of
an anisotropic plasma-scale
turbulent magnetic field and an optically thin limit by using the standard
calculation formulae of relativistic blast wave dynamics and synchrotron emission. We have clarified
that since GRB afterglows are observed as point sources, the net linear polarizations are determined
by the image shape intensity which depends on the observed frequency and its asymmetry due to the
collimation of outflow (see Figure~\ref{fig:images}). Our calculation results have shown that the PD
at $\nu < \nu_T$ is always lower than the optical PD (Figures~\ref{fig:polcurve} and \ref{fig:pinu}),
since the region with low polarized intensity in the image is larger at $\nu < \nu_T$, i.e., $\tilde{S}_Q
/\tilde{S}_I$ is smaller at $\nu < \nu_T$ (see Figure~\ref{fig:images} and \ref{fig:contrast}). We find
that the PD gradually varies above $\nu \sim \nu_T$, and the break frequency of the high PD regime to
the low one is $\nu_B \sim 40 \nu_T$ at $T < T_{\rm j}$, while $\nu_B/\nu_T$ rapidly decreases at $T >
T_{\rm j}$ (Figure~\ref{fig:pinu}). We also show that the difference in PAs between the high and low PD
regimes is zero or 90 degrees. Thus, the simultaneous polarimetric observations of late-phase GRB afterglows
at the radio (typically $\nu < \nu_B$) and optical bands (typically $\nu > \nu_B$) would be a new
determinative
test of the anisotropic plasma-scale turbulent magnetic field model.
\par
Radio polarizations have been measured recently for GRB 171205A \citep{urata19} and GRB 190114C
\citep{laskar19} with ALMA at the frequency of $\nu_{\rm ALMA} \simeq 97.5\;$GHz, which
is typically higher
than the synchrotron self-absorption frequency. For GRB 171205A, the polarization of forward shock
emission was measured as PD $\simeq  0.27 \pm 0.04\;\%$. This is significantly lower than the typical
optical PD of late-phase afterglows of
$\sim 1-3\;\%$ \citep{covino04}. The observed intensity light curve and spectrum indicate
$\nu_T \sim 150\;$GHz and $T \sim 2.5 T_{\rm j}$ at the polarization measurement \citep{urata19}, so
that we find $\nu_{\rm ALMA} < \nu_B$. These observational indications appear consistent with the
anisotropic plasma-scale turbulent field model (although we have no simultaneous optical polarization measurement
for this GRB). However, the observed PAs seem to vary in the range of $90.5-104.5\;$GHz, which are at
odds with the model. Interestingly, for GRB 190114C, the radio polarimetric observations performed
{\it for the reverse shock emission} show gradual temporal change of PA, which rules out the plasma-scale
turbulent field model \citep{laskar19}. 
\par
The magnetic field structure of GRB forward shock blast waves has been examined by using only optical
polarimetric data so far \citep{covino16}. The temporal PA flips by 90 degrees, predicted by the plasma-scale
field model, were observed in some GRBs such as GRB 091018 and GRB 121024A \citep{wiersema14} while not
observed in some other GRBs such as GRB 020813 \citep{lazzati04}. Here,
we should note that the temporal PA
flips are not necessarily observed for jets with angular structure \citep{rossi04} and for the shocked region
including ordered field \citep{granot03}. The angular structure of GRB jets may be constrained by modeling
the densely observed intensity light curves, as performed for short GRB 170817A with
the detection of the gravitational
wave of the progenitor system
\citep[e.g.][]{gill20,takahashi19}. The afterglow of GRB 091208B observed by the Kanata
telescope at $T = 149-706\;$s appears to be from the forward shock and has
a PD of $10.4\pm2.5\;\%$. Such a high PD
at the early phase does not favor the plasma-scale turbulent field model \citep{uehara12}.
\par
The magnetic field in the blast waves may not be dominated by the turbulent component on
the plasma skin-depth
scale but by that on the hydrodynamic scale
\citep{sironi07,inoue13,duffell14,tomita19}. Such field structure
is suggested in observed non- (or sub-) relativistically expanding blast waves, for example, Tycho's supernova
remnant \citep{shimoda18} and AT2018cow \citep{huang19}. Especially for the former example, a Kolmogorov scaling
is seen in the magnetic field energy spectrum. The differences between the optical and radio polarizations in
this field structure should be clarified in separate papers. Probably the temporal variation (or the distribution
in the egg) of the field coherence scale may be crucial.
\par
Furthermore, for the hydrodynamic-scale magnetic field case, the Faraday rotation effects within the shocked
region should be taken into account. If only a fraction $f_{\rm e}$ of the electrons swept by the shock are energized
to form the nonthermal energy spectrum at $\gamma_{\rm e} \ge \gamma_{\rm m} \sim \Gamma \epsilon_{\rm e} m_{\rm p}/
m_{\rm e}$, the Faraday depolarization by non-energized thermal electrons at $\gamma_{\rm e} \sim \Gamma$ may
suppress the PD even at $\nu > \nu_{\rm a}$ \citep[in the
ALMA band,][]{toma08,urata19} and the PAs may be a complex function of
$\nu$ \citep{sokoloff98}. In the case of $f_{\rm e} < 1$, $E_{\rm iso}$ should be $1/f_{\rm e}$ times larger than the ordinary
estimates under the assumption of $f_{\rm e}=1$ \citep{eichler05}. While some particle-in-cell simulations suggest
$f_{\rm e} \simeq 1$ \citep{sironi11,kumar_eichler15}, it should be confirmed by observations \citep[see also][]{ressler17,warren18,asano20}.
Simultaneous polarimetric observations of GRB afterglows at the optical and multiple radio bands with ALMA
and more theoretical investigations offer the
potential power for understanding the magnetic field structure
and the electron energy distribution in the downstream of relativistic collisionless shocks.

\acknowledgments
We thank J. Matsumoto, S. Tomita, Y. Urata, and A. Kuwata for useful discussions.
This work is partly supported by JSPS Grants-in-Aid for Scientific Research No. 18H01245 (KT)
and No. 20J01086 (JS).

%






\bibliography{apj_st19}{}
\bibliographystyle{aasjournal}

\end{document}